\documentstyle[preprint,aps,   epsf]{revtex}
\tightenlines
\begin{document}
\title{Plateaux Transitions from S-matrices based on $SL(2,Z)$ Invariant
Field Theories}
\author{I. Devetak and A. LeClair}
\address{Newman Laboratory, Cornell University, Ithaca, NY
14853.}
\date{\today}
\maketitle

\begin{abstract}

A scattering description is proposed for a boundary perturbation 
of a $c=1$ $SL(2,Z)$ invariant conformal field theory.  The
bulk massless S-matrices are of the form of Zamolodchikov's
staircase model.  Using the boundary version of the thermodynamic
Bethe ansatz, we show that the boundary free energy  goes through 
a series of integer valued plateaux as a function of system size. 

\end{abstract}
\vskip 0.2cm
\pacs{PACS numbers: }
\narrowtext
%
%
%
%
\def\oti{{\otimes}}
\def\bra#1{{\langle #1 |  }}
\def\lb{ \left[ }
\def\rb{ \right]  }
\def\tilde{\widetilde}
\def\bar{\overline}
\def\hat{\widehat}
\def\*{\star}
\def\[{\left[}
\def\]{\right]}
\def\({\left(}		\def\BL{\Bigr(}
\def\){\right)}		\def\BR{\Bigr)}
	\def\BBL{\lb}
	\def\BBR{\rb}
%
%
\def\zb{{\bar{z} }}
\def\zbar{{\bar{z} }}
\def\frac#1#2{{#1 \over #2}}
\def\inv#1{{1 \over #1}}
\def\half{{1 \over 2}}
\def\d{\partial}
\def\der#1{{\partial \over \partial #1}}
\def\dd#1#2{{\partial #1 \over \partial #2}}
\def\vev#1{\langle #1 \rangle}
\def\ket#1{ | #1 \rangle}
\def\rvac{\hbox{$\vert 0\rangle$}}
\def\lvac{\hbox{$\langle 0 \vert $}}
\def\2pi{\hbox{$2\pi i$}}
\def\e#1{{\rm e}^{^{\textstyle #1}}}
\def\grad#1{\,\nabla\!_{{#1}}\,}
\def\dsl{\raise.15ex\hbox{/}\kern-.57em\partial}
\def\Dsl{\,\raise.15ex\hbox{/}\mkern-.13.5mu D}
%
%
\def\th{\theta}		\def\Th{\Theta}
\def\ga{\gamma}		\def\Ga{\Gamma}
\def\be{\beta}
\def\al{\alpha}
\def\ep{\epsilon}
\def\vep{\varepsilon}
\def\la{\lambda}	\def\La{\Lambda}
\def\de{\delta}		\def\De{\Delta}
\def\om{\omega}		\def\Om{\Omega}
\def\sig{\sigma}	\def\Sig{\Sigma}
\def\vphi{\varphi}
%
%
\def\CA{{\cal A}}	\def\CB{{\cal B}}	\def\CC{{\cal C}}
\def\CD{{\cal D}}	\def\CE{{\cal E}}	\def\CF{{\cal F}}
\def\CG{{\cal G}}	\def\CH{{\cal H}}	\def\CI{{\cal J}}
\def\CJ{{\cal J}}	\def\CK{{\cal K}}	\def\CL{{\cal L}}
\def\CM{{\cal M}}	\def\CN{{\cal N}}	\def\CO{{\cal O}}
\def\CP{{\cal P}}	\def\CQ{{\cal Q}}	\def\CR{{\cal R}}
\def\CS{{\cal S}}	\def\CT{{\cal T}}	\def\CU{{\cal U}}
\def\CV{{\cal V}}	\def\CW{{\cal W}}	\def\CX{{\cal X}}
\def\CY{{\cal Y}}	\def\CZ{{\cal Z}}

\def\rvac{\hbox{$\vert 0\rangle$}}
\def\lvac{\hbox{$\langle 0 \vert $}}
\def\comm#1#2{ \BBL\ #1\ ,\ #2 \BBR }
\def\2pi{\hbox{$2\pi i$}}
\def\e#1{{\rm e}^{^{\textstyle #1}}}
\def\grad#1{\,\nabla\!_{{#1}}\,}
\def\dsl{\raise.15ex\hbox{/}\kern-.57em\partial}
\def\Dsl{\,\raise.15ex\hbox{/}\mkern-.13.5mu D}
%
%
%
\font\numbers=cmss12
\font\upright=cmu10 scaled\magstep1
\def\stroke{\vrule height8pt width0.4pt depth-0.1pt}
\def\topfleck{\vrule height8pt width0.5pt depth-5.9pt}
\def\botfleck{\vrule height2pt width0.5pt depth0.1pt}
\def\Zmath{\vcenter{\hbox{\numbers\rlap{\rlap{Z}\kern
0.8pt\topfleck}\kern 2.2pt
                   \rlap Z\kern 6pt\botfleck\kern 1pt}}}
\def\Qmath{\vcenter{\hbox{\upright\rlap{\rlap{Q}\kern
                   3.8pt\stroke}\phantom{Q}}}}
\def\Nmath{\vcenter{\hbox{\upright\rlap{I}\kern 1.7pt N}}}
\def\Cmath{\vcenter{\hbox{\upright\rlap{\rlap{C}\kern
                   3.8pt\stroke}\phantom{C}}}}
\def\Rmath{\vcenter{\hbox{\upright\rlap{I}\kern 1.7pt R}}}
\def\Z{\ifmmode\Zmath\else$\Zmath$\fi}
\def\Q{\ifmmode\Qmath\else$\Qmath$\fi}
\def\N{\ifmmode\Nmath\else$\Nmath$\fi}
\def\C{\ifmmode\Cmath\else$\Cmath$\fi}
\def\R{\ifmmode\Rmath\else$\Rmath$\fi}






\section{Introduction}

A simple model was recently proposed in connection with Quantum Hall
plateaux transitions\cite{andre}.  After bosonization, the model consists
of a scalar field coupled to a pure gauge field with a boundary
interaction incorporating a circular defect line of impurities.  
The $c=1$ conformal field theory possesses an $SL(2,\Zmath)$ symmetry.
Though the precise connection with the complete Landau problem in 
the presence of disorder is not entirely clear, the model does have
some promising features, and we continue to investigate it in this paper. 

In the next section an exact S-matrix description of the theory is proposed. 
We cannot claim to give here an absolute   derivation of this S-matrix 
since we are missing a more precise treatment of the zero-mode 
constraint (Eq. (\ref{2.3})).  Rather we give a suggestive derivation 
based on the prescriptive treatment described in \cite{andre}.  

The proposed scattering description allows an exact computation of the
boundary (impurity) contribution to the free energy, $\log \, g$. 
 It reveals an
infinite series of plateaux at integer values of $g$.

\section{S-matrices}

\def\vphi{\varphi}
\def\vphid{\tilde{\vphi}}
\def\beq{\begin{equation}}
\def\eeq{\end{equation}}
\def\ghat{\hat{g}}

Let us first summarize some of the features of the model
 described  in \cite{andre}. 
The conformal field theory  was defined by
the euclidean action 
\beq
\label{2.1}
S = \int dt  d\sigma 
\( \inv{8\pi} \( \d_\mu \vphi \)^2 
- \frac{i}{2\pi \ghat} \ep_{\mu\nu} \d_\nu \vphi A_\mu 
- \inv{2\pi \ghat^2} A_\mu^2 \) 
\eeq
where $A_\mu$ is the electro-magnetic gauge field.  
The above theory lives on a cylinder with $0<\sigma <2\pi$,
and in the folded boundary version which we consider here, $0<t<\infty$. 
The lagrangian possesses the gauge symmetry
\beq
\label{2.2}
\CL_{\rm cft} (\vphid + \frac{2}{\ghat} \lambda , 
A_\mu + \d_\mu \lambda ) = 
\CL_{\rm cft} (\vphid, A_\mu ) 
\eeq
where $\d_\mu \vphid = -i \ep_{\mu\nu} \d_\nu \vphi$.  
The gauge field was taken to be a pure, singular gauge $A_\mu = \d_\mu \chi$,
and the theory was supplemented by the zero mode constraint
\beq
\label{2.3}
\frac{\theta}{2\pi \ghat} \oint dx_\mu \d_\mu \vphi = 
\oint dx_\mu \d_\mu \chi 
\eeq
When $\sigma_{xx} = 0$, the parameter $\theta$ was related to the
Hall conductivity as $\sigma_{xy} = 1/\theta$.  

The pure gauge field can be gauged away using 
\beq
\label{2.4}
\CL_{\rm cft} (\vphid + \frac{2}{\ghat} \chi , \d_\mu \chi ) 
= \CL_{\rm cft} (\vphid , 0) 
\eeq
Using the zero mode constraint Eq. (\ref{2.3}),
 this leads us to consider  the gauge transformation 
\beq
\label{2.5}
\vphid \to \vphid + \frac{\theta}{\pi \ghat^2} \, \vphi
\eeq
From the transformation (\ref{2.5}) on the primary fields of the
theory, one finds that the partition function has an 
$SL(2, \Zmath)$ invariance acting on the modular parameter 
$\tau = \theta/2\pi + i \ghat^2 /2$.  

Upon adding a circular defect line of impurities, in the folded
theory this corresponds to a boundary term in the action 
$\int d\sigma \cos \( \vphid (0,\sigma)/\sqrt{2} \)$, where the
boundary is at $t=0$.  
Performing the gauge transformation (\ref{2.5}) led to the boundary
field theory 
\beq
\label{2.6}
S = S_{bcft} + 
\lambda \int d\sigma ~ 
\cos 
\( \frac{b}{2} \( \vphid + \frac{\theta}{\pi \ghat^2} \vphi \)  \) 
\eeq
where $b= \sqrt{2}$.  The boundary conformal field theory
$S_{bcft}$ is that of a free scalar field, but with the unusual
boundary condition\cite{andre}:
\beq
\label{2.7}
\d_t \vphi - \frac{i\theta}{\pi \ghat^2} \d_\sigma \vphi = 0 , ~~~~~~
(t=0)
\eeq

The above theory is massless in the bulk, but with a mass/length
scale at the boundary.  To obtain an S-matrix description of this
theory we follow the ideology described in \cite{Saleur}.  Namely,
we imagine turning on an integrable bulk interaction such that
the integrability is not destroyed by the boundary interactions.
This selects a special basis of particles in the bulk that diagonalize
the boundary interactions.  We then take a massless limit in the bulk.
A bulk interaction that is compatible with the boundary interaction,
as far as integrability goes, is 
$\delta S_{\rm bulk} = \Lambda \int dt d\sigma ~ \cos (b\vphid)$,
where again $b=\sqrt{2}$.   
The reason is, that since $(\d_\mu \vphi)^2 = - (\d_\mu \vphid )^2$,
written in terms of $\vphid$,  the above theory, without the gauge
field,  is equivalent to the boundary
sine-Gordon model, which is known to be integrable\cite{ghoshal}.  
Performing the same gauge transformation Eq. (\ref{2.5}) leads us
to consider $\delta S_{\rm bulk} = \Lambda \int dt d\sigma 
\cos b ( \vphid + \theta \vphi/ \pi \ghat^2 )$.  

\def\thetah{\hat{\theta}}

Next
consider the boundary condition Eq. (\ref{2.7}).  Using 
$\ep_{\sigma t} = - \ep_{t\sigma} = 1$, one has $i\d_\sigma 
\vphi = \d_t \vphid$.  Thus the boundary condition can be written
as 
\beq
\label{2.8}
\d_t \( \vphi - \frac{\theta}{\pi \ghat^2} \vphid \) = 0 , ~~~~~(t=0)
\eeq
This is a Neumann boundary condition for the combination 
$\vphi - \theta\vphid / \pi \ghat^2$. 
All of this leads us to consider the bulk theory
\beq
\label{2.8b}
S_{\rm bulk} = \int dt d\sigma \[ 
\inv{8\pi} \( \d_\mu \( \vphi - \frac{\thetah}{2\pi} \vphid \) \)^2 
+ \Lambda \cos \sqrt{2} \( \vphid + \frac{\thetah}{2\pi} \vphi \) \] 
\eeq
where we have defined $\thetah = 2 \theta /\ghat^2 $.  
When $\Lambda = 0$, the boundary version of the above free theory leads
to the boundary condition (\ref{2.8}). 

 Let us now rewrite the theory in terms of $\vphid$.
 As far as the bulk theory
is concerned, we can drop the topological term $\d_\mu \vphi 
\d_\mu \vphid$, since it is identically zero. 
 Using $(\d_\mu \vphid)^2 = - (\d_\mu \vphi )^2$,
and defining a rescaled field 
$\vphid \to i \vphid/\sqrt{1-(\thetah/2\pi)^2} $,
one finds 
\beq
\label{dualS}
S_{\rm bulk} = \int dt d\sigma \(
\inv{8\pi} (\d_\mu \vphid)^2 + \Lambda \cosh
\( b_L \vphid_L + b_R
\vphid_R \) \)
\eeq
where we  have used the left-right decompositions $\vphi = \vphi_L + 
\vphi_R$, $
\vphid = \vphi_L - \vphi_R$, and 
\beq
\label{2.10}
\frac{b_L}{\sqrt{2}} =  \frac{1+ \thetah/2\pi}{\sqrt{1-(\thetah/2\pi)^2} } , 
~~~~~
\frac{b_R}{\sqrt{2}} =  \frac{1- \thetah/2\pi}{\sqrt{1-(\thetah/2\pi)^2} } , 
\eeq

When $b_L = b_R$, 
the above bulk theory is the well-known sinh-Gordon model.  
Consider now the massless limit $\Lambda \to 0$.  The theory
has both left and right moving particles.  For the right-movers
we parameterize the energy and momentum by $E = P = \mu e^\beta$,
and for the left movers $E=-P = \mu e^{-\beta}$, where $\beta$ is
a rapidity and $\mu$ an arbitrary energy scale.  One can describe
the theory in terms of an S-matrix $S_{LL} (\beta)$ for the left-movers
and $S_{RR} (\beta)$ for the right-movers\cite{ZamoZamo}. 
From the form of the bulk interaction in Eq. (\ref{dualS}) it is 
clear that $S_{LL}$ is the sinh-Gordon S-matrix defined by the
sinh-Gordon coupling $b_L$, whereas $S_{RR}$ is defined by 
$b_R$.  One argument is the following.  In the sine-Gordon version,
one can characterize the S-matrices by the non-local quantum affine
symmetry\cite{BL}.  This symmetry survives in the massless limit,
and the left-moving (right-moving) quantum affine symmetry will have
$q$-deformation parameter defined by $b_L$ $(b_R)$.  Using the  known
S-matrix for the sinh-Gordon model\cite{Korepin},   the
result is thus 
\beq
\label{2.13} 
S_{LL} (\beta) = \frac{ \tanh \inv{2} (\beta - i \pi \gamma_L )}
{\tanh \inv{2} (\beta + i \pi \gamma_L )} 
\eeq
where 
\beq
\label{2.14}
\gamma_L = \frac{ b_L^2 }{2 + b_L^2 } = \inv{2} + \frac{\thetah}{4\pi}  
\eeq
Similarly, $S_{RR}$ is given by Eq. (\ref{2.13}) with 
\beq
\label{gammar}
\gamma_R = \frac{b_R^2}{2 + b_R^2 } = \inv{2} - \frac{\thetah}{4\pi}
 = 1- \gamma_L 
\eeq
 Using
the invariance of the S-matrix under $\gamma \to 1-\gamma$, one sees
that $S_{LL} = S_{RR}$.  

We  discuss now the $SL(2,\Zmath)$ properties of the above S-matrix.
As described in \cite{andre}, the bulk conformal field theory has
an $SL(2,\Zmath)$ invariance acting on the modular parameter 
$\tau = \theta/2\pi + i \ghat^2 /2 $.  More specifically, the partition
function on the torus is invariant under $SL(2,\Zmath)$.  Since
$S_{LL}, S_{RR}$ provide a scattering description of this conformal field
theory,    we expect the S-matrix to be at least in part characterized
by this symmetry.   
Under $\tau \to -1/\tau$, one has that $(\ghat, \theta) 
\to (\ghat' , \theta')$ where
\beq
\label{Z5}
\ghat'^2 =  \frac{\ghat^2}{\( \ghat^4/4  + (\theta/2\pi )^2  \)}
, ~~~~~~
\theta' = - \frac{\theta} { \(   \ghat^4/4  + (\theta/2\pi )^2 \) }
\eeq
Note that under this transformation, $\theta/\ghat^2 
\to - \theta/\ghat^2$, which implies  $\thetah \to - \thetah$.  Thus
$\tau \to - 1/\tau$ simply exchanges left and right movers:  
\beq
\label{sl4}
\tau \to - 1/\tau  ~~ 
\Longrightarrow ~~
b_L^2  \to b_R^2 
\eeq
Since $S_{LL} = S_{RR}$, the scattering theory has this invariance. 

The other independent generator of $SL(2,\Zmath)$ corresponds to 
$\tau \to \tau + 1$, which corresponds to $\theta \to \theta + 2\pi$. 
The S-matrices turn out to only be invariant under a multiple of this 
transformation.
Namely, using the invariance of the S-matrices under $\gamma \to 
\gamma + 2$, one verifies their invariance under 
$\theta \to \theta + 4\pi \ghat^2$.  For the original fermion model
considered in \cite{andre} with $
\ghat = \sqrt{2}$ this corresponds to $\theta \to \theta + 8\pi$.   
The significance of this is not entirely clear.

We now consider performing the analytic continuation $\theta \to 
i \theta$.  There are at least two justifications for doing this. 
First, the Hall conductivity  computed in \cite{andre} is 
imaginary unless one performs this continuation.   Second, 
in order to formulate the above scattering description one needs
to have identified a ``time'' by continuing to Minkowski space.   
Let us analytically continue to Minkowski
space by identifying $\sigma$ as the time (based on the boundary
interaction) and letting $\sigma \to i \sigma$.  The Minkowski action
$S_M$ is obtained from the euclidean one by $S_M = i S (\sigma \to i\sigma)$.
This leads to the analytic continuation of
$\theta \to i \theta$. 
  To see this, let us rescale $\chi \to
\theta \chi$.  Then the euclidean topological term is
\beq
\label{2.11}
S^{\rm top} \propto i \theta \int dt d\sigma
(\d_t \vphi \d_\sigma \chi - \d_\sigma \vphi \d_t \chi )
\eeq
Letting $\sigma \to i \sigma$ and identifying $S_M$ as above one finds
that the Minkowski action is obtained by the analytic continuation
$\theta \to i \theta$.  Performing this continuation in Eq. (\ref{2.10})
one finds that $b_R = b_L^*$, $b_L b_R = 2$ and 
that the S-matrices now have the parameters:
\beq
\label{gamcon}
\gamma_L = \inv{2} + i \frac{\thetah}{4\pi}, ~~~~~
\gamma_R = \gamma_L^* 
\eeq

The resulting S-matrix has the same form as the bulk massive staircase
model of Zamolodchikov \cite{Alyosha}, where there $\gamma$ was
taken as $\gamma_\pm = 1/2 \pm i \theta_0 /\pi $.  However since
we are here dealing with a massless bulk theory, the interpretation
is different and in fact our theory doesn't suffer from some of
the problems of interpretation of the bulk massive staircase model.  
Namely, since there is no  bulk interaction in the massless
limit, just a free scalar field, the theory doesn't have the
reality problems of the massive case arising from complex values of
the sinh-Gordon coupling $b$.   
The above scattering theory 
is a limiting case of the bulk massive theories studied in 
\cite{Saleur3}.

Finally, the interactions at the boundary are described by a reflection 
S-matrix $R(\beta)$ for reflection of the above particles off the
boundary.  In the massless sinh-Gordon model, this reflection
S-matrix, which can be obtained by taking the explicit massless
limit of the boundary sinh-Gordon model, is known to be independent
of the sinh-Gordon coupling $b_L$.  (See \cite{Saleur2}).  
The result should be the same in our case and we thus take
\beq
\label{2.16} 
R(\beta) = \tanh (\beta/2 - i\pi/4)
\eeq
The physical S-matrix for right  movers is 
$R(\beta - \beta_B)$, and for left movers $R(\beta + \beta_B)$, where
$\mu e^{\beta_B}$ is defined as a physical boundary energy scale.  

\bigskip

\section{Plateaux Transitions in the Boundary Entropy} 

\def\vep{\varepsilon}

We consider now the theory on a semi-infinite cylinder of circumference
$L$, where $\sigma$ runs along the circumference and 
$\sigma \sim \sigma + L$ and as before $0<t<\infty$.  Viewing
$\sigma$ as the time, the Hilbert space lives on the semi-infinite
line $0<t<\infty$, and finite size $L$ effects can be computed from
the ``L-channel'' thermodynamic Bethe ansatz\cite{Fendley}\cite{Mussardo}. 
Of interest is the boundary entropy $\log \, g$\cite{Affleck}, defined
as the contribution to the free energy that is independent of the length
$R$ of the cylinder, $\log Z = \log g + \log Z_{\rm bulk}$, where
$\log Z_{\rm bulk} $ is proportional to $R$.  
Viewing the theory as a $1+1$ dimensional quantum system, $g$
represents the ground state degeneracy, i.e. it counts the number of states
at the boundary.   

Using the formulas in \cite{Mussardo}, one has
\beq
\label{3.1}
\log g = \inv{2\pi i} \int_{-\infty}^\infty d\beta 
\[ \d_\beta \log R (\beta)\] ~ \log \( 1+ e^{-\vep (\beta)} \)  
\eeq
where $\vep (\beta)$ is a solution of the integral equation 
\beq
\label{3.2}
\vep (\beta) = \frac{L}{2\xi_B} e^\beta 
- \inv{2\pi i} 
 \int d\beta' ~ \[ \d_\beta \log S_{LL}  (\beta - \beta') \]  
\log \( 1 + e^{-\vep (\beta')} \) 
\eeq
In the above equation, $\xi_B$ is a boundary length scale 
$1/\xi_B = \mu e^{\beta_B}$; it defines an energy scale at
the boundary $E_B = 1/\xi_B$.  

The numerical solution of $g$ as a function of $\xi_B/L$ is shown 
for the two values of $\thetah = 200, 100$ in figures 1 and 2.  
There are two important features of these figures.  The first is 
that on the plateaux $g$ takes on the series of integers $g=1,2,3,..$,
as anticipated in \cite{Saleur3}.
The second is that the plateaux are more clearly defined as 
$\thetah$ is increased.

\section{Discussion}

We conclude with a discussion of the possible implications of our results
for the proposal made in \cite{andre} in connection with the Quantum
Hall transitions.  The integer valued plateaux that we found in the
boundary state degeneracy is a promising feature.  The meaning
of the boundary entropy suggests that as the scale $L$ is 
decreased, the model goes through a series of transitions where at
each transition one more state becomes localized at the impurities. 
In order to verify this picture one needs to relate the boundary
entropy of our 2-dimensional model to the $2+1$ dimensional system. 

The significance of $\thetah = \pi$, as discussed in \cite{andre}, 
cannot be seen from what we have done in this paper since we have not
studied the conductivities $\sigma_{xx}, \sigma_{xy}$.  In 
the conformal field theory it was argued that $
\thetah = \pi$ implies $\sigma_{xx} = 0$, $\sigma_{xy} = 1/\theta$.  One
needs to study the conductivities  in the presence of the boundary
interaction.  It should be possible to do this using form factors. 
Since the conductivities are reduced to boundary effects, it seems likely
that the transitions in the boundary entropy will entail transitions in
the conductivity, but only a detailed analysis will tell.  

A related issue which needs clarification concerns whether the scattering
theory described here implicitly treats the boundary condition in a way
that is consistent with the perturbative treatment in \cite{andre}, which,
when $\thetah = \pi$, led to the correlation length exponent $20/9$.


\vbox{
\vskip 1.5 truein
\epsfysize=16cm
\epsfxsize=16cm
\epsffile{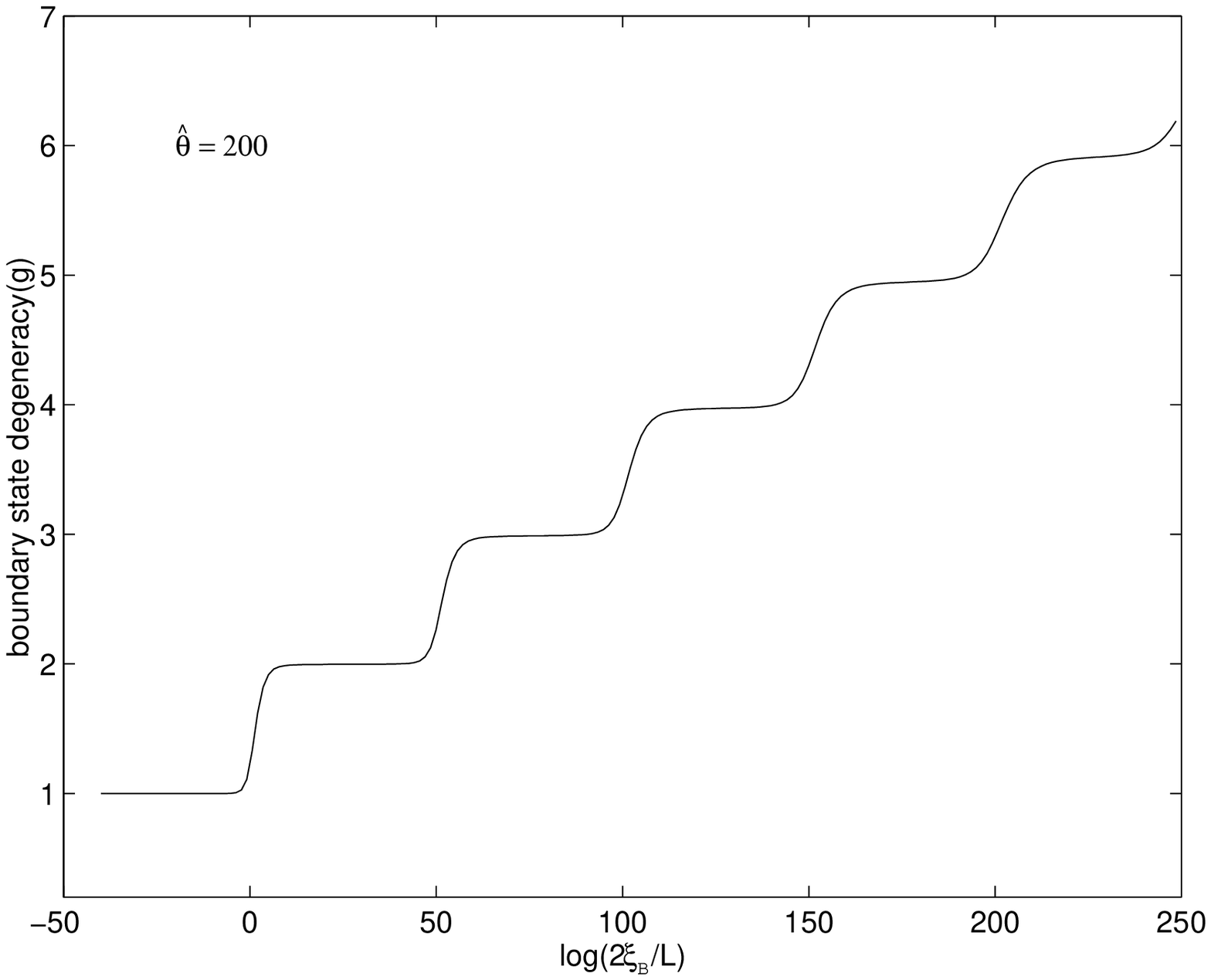}
\begin{figure}
\caption[]{Ground state degeneracy as a function of $L/\xi_B$ for
$\thetah = 200$.}
\end{figure}
}

\vbox{
\vskip 1.5 truein
\epsfysize=14cm
\epsfxsize=14cm
\epsffile{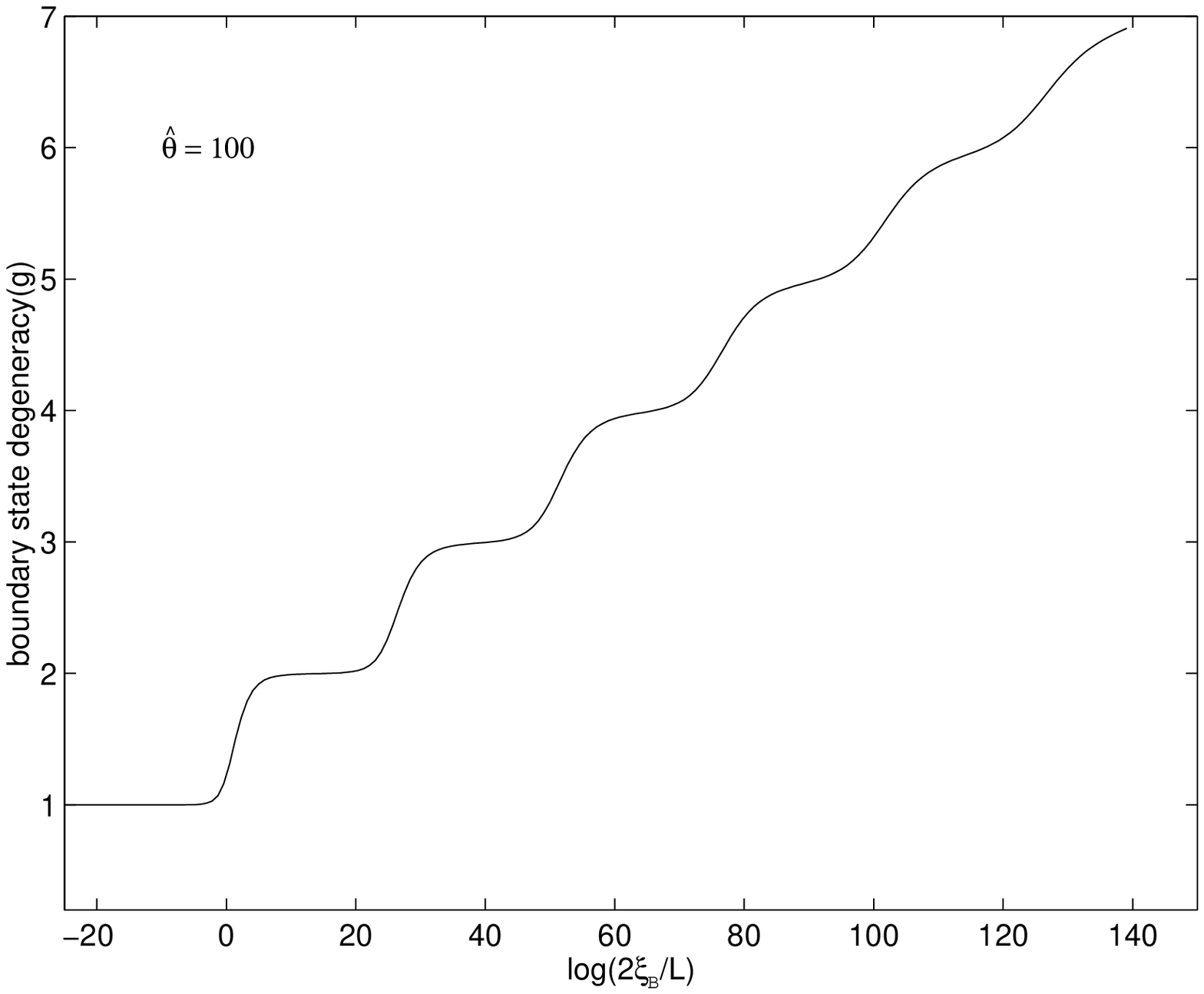}
\begin{figure}
\caption[]{Ground state degeneracy as a function of $L/\xi_B$
for $\thetah = 100$.}
\end{figure}
}


\begin{references}

\bibitem{andre}  A. LeClair, {\it Gauge Invariance and the Critical
Properties of Quantum Hall Plateaux Transitions}, cond-mat/9904414.

\bibitem{Saleur}  P. Fendley, H. Saleur and N. P. Warner, Nucl. Phys.
B 430 (1994) 577. 
 

\bibitem{ghoshal}  S. Ghoshal and A. B. Zamolodchikov, Int. J. Mod. Phys.
A9 (1994) 3841. 

\bibitem{ZamoZamo}  A. B. Zamolodchikov and Al. B. 
Zamolodchikov, Nucl. Phys. B379 (1992) 602. 

\bibitem{BL}  D. Bernard and A. LeClair, Commun. Math. Phys. 142 (1991) 99.

\bibitem{Korepin}  I. Ya. Arefyeva and V. E. Korepin, Pisma v ZhETF 20
(1974) 680; B. Schroer, T. T. Truong and P. H. Weisz, Phys. Lett. 63B
(1976) 422. 


\bibitem{Alyosha}  Al. B. Zamolodchikov, {\it Resonance Factorized
Scattering and Roaming Trajectories}, Ecole Normale Sup\'erieure
preprint, ENS-LPS-335, 1991, otherwise unpublished.
  
\bibitem{Saleur3}  F. Lesage, H. Saleur and P. Simonetti,
Phys. Lett. B427 (1998) 85.

\bibitem{Saleur2}  F. Lesage, H. Saleur and S. Skorik, 
Nucl.Phys. B474 (1996) 602.  


\bibitem{Fendley}  P. Fendley and H. Saleur, Nucl. Phys. B428 (1994) 681. 

\bibitem{Mussardo} A. LeClair, G. Mussardo, H. Saleur and S. Skorik, 
Nucl. Phys. B 453 (1995) 581. 

\bibitem{Affleck} I. Affleck and A. W. W. Ludwig, Phys. Rev. Lett. 67
(1991) 161. 


\end{references}
\end{document}